\documentclass[twocolumn]{aastex631}
\usepackage{amsmath}
\usepackage{hyperref}
\usepackage[capitalise,nameinlink,noabbrev]{cleveref}

\usepackage[nolist]{acronym}

\crefname{section}{Section}{Sections}
\Crefname{section}{Section}{Sections}

\newcommand{\unit}[1]{\ensuremath{\, \mathrm{#1}}}

\graphicspath{{./}{figures/}}

\received{XXX}
\revised{XXX}
\accepted{XXX}

\shorttitle{The VSI in Stratified Disks}
\shortauthors{Yun et al.}

\begin{document}
\title{Vertical Shear Instability in Thermally-Stratified Protoplanetary Disks: I. A Linear Stability Analysis}

\author[0000-0003-4353-294X]{Han-Gyeol Yun}
\affiliation{Department of Physics \& Astronomy, Seoul National University, Seoul 08826, Korea}
\affiliation{SNU Astronomy Research Center, Seoul National University, 1 Gwanak-ro, Gwanak-gu, Seoul 08826, Republic of Korea}

\author[0000-0003-4625-229X]{Woong-Tae Kim}
\affiliation{Department of Physics \& Astronomy, Seoul National University, Seoul 08826, Korea}
\affiliation{SNU Astronomy Research Center, Seoul National University, 1 Gwanak-ro, Gwanak-gu, Seoul 08826, Republic of Korea}

\author[0000-0001-7258-770X]{Jaehan Bae}
\affiliation{Department of Astronomy, University of Florida, Gainesville, FL 32611, USA}

\author[0000-0002-2641-9964]{Cheongho Han}
\affiliation{Department of Physics, Chungbuk National University, Cheongju 28644, Republic of Korea}

\email{hangyeol@snu.ac.kr, wkim@astro.snu.ac.kr, jbae@ufl.edu, cheongho@astroph.chungbuk.ac.kr}

\begin{acronym}
  \acro{PPD}{protoplanetary disk}
  \acro{ALMA}{Atacama Large Millimeter/submillimeter Array}
  \acrodefplural{PPD}{protoplanetary disks}
  \acro{MRI}{magnetorotational instability}
  \acro{MHD}{magnetohydrodynamic}
  \acro{VSI}{vertical shear instability}
  \acro{FWHM}{full-width at half maximum}
\end{acronym}

\begin{abstract}
Vertical shear instability (VSI), driven by a vertical gradient of rotational angular velocity, is a promising source of turbulence in protoplanetary disks. We examine the semi-global stability of thermally stratified disks and find that the VSI consists of surface and body modes: surface modes are confined to regions of strong shear, while body modes extend perturbations across the disk, consistent with the previous findings.  In thermally stratified disks, surface modes bifurcate into two branches. The branch associated with the strongest shear at mid-height exhibits a higher growth rate compared to the branch near the surfaces. Surface modes generally grow rapidly and require a high radial wave number $k_R$, whereas body mode growth rates increase as $k_R$ decreases. Thermal stratification enhances the growth rates of both surface and body modes and boosts VSI-driven radial kinetic energy relative to vertical energy. Our results suggest that simulations will initially favor surface modes with large $k_R$, followed by an increase in body modes with smaller $k_R$, with faster progression in more thermal stratified disks.
\end{abstract}


\section{Introduction} \label{sec:intro}

A \ac{PPD} is a type of accretion disks through which gas moves radially inward slowly toward a central star.  Early works of \citet{ss1973} and \citet{lp1974} have shown that the radial inflow of gas can be driven by  turbulent viscosity responsible for an outward transport of angular momentum. Turbulence can also control the dust distribution and particle growth, greatly affecting the formation of planets \citep{lu2019}. Therefore, the study of turbulence inside \acp{PPD} is vital to the understanding of the disk evolution and the planet formation processes.

It has long been believed that turbulence inside \acp{PPD} is driven by \ac{MRI} which operates when magnetic fields are very weak and the gas is sufficiently ionized, as in hot accretion disks surrounding compact objects such as white dwarfs and black holes (see, e.g., \citealt{bh1991, bb1994}). However, the low level of ionization due to strong shielding and low gas temperature in \acp{PPD} provide a hostile condition for the operation of \ac{MRI} \citep{bb1994, g1996}, forming \ac{MRI}-dead zones in non-ideal \ac{MHD} simulations \citep{bs2013, sb2013, lk2014}. Since the \ac{MRI}-dead zones can occupy a significant portion of a \ac{PPD} \citep{tf2014}, it is necessary to find alternative mechanisms that drive turbulence and angular momentum transport in \acp{PPD} \citep{lu2019}. 

One possible attractive mechanism is \ac{VSI} which is a pure hydrodynamic instability. It was originally proposed in the context of gas mixing in differentially rotating stars \citep{gs1967, f1968}, and has been applied to turbulence generation in \acp{PPD} both analytically \citep{ub1998, u2003} and numerically (e.g., \citealt{ng2013}). In a \ac{PPD}, the \ac{VSI} operates when there is a non-vanishing gradient of the rotational angular velocity along the vertical direction within the disk \citep{u2003, ng2013}. Since the vertical shear in the rotational angular velocity can readily be induced by the radial temperature gradient of the disk \citep{ub1998, ng2013}, all \acp{PPD} are unstable to the \ac{VSI} for practical purposes. Numerical simulations show that the \ac{VSI} produces a meridional circulation pattern, a radially narrow and vertically elongated, axisymmetric structure \citep{ng2013, sk2014}, eventually developing anisotropic turbulence with stronger vertical motions than in the radial direction \citep{sk2017}. These perturbations can further evolve into non-axisymmetric flows such as vortices -- a typical feature observed in full three-dimensional \ac{VSI} simulations -- that can persist for several hundred orbits \citep{rn2016, mk2018, mk2020, ft2020}. Although vertical buoyancy and magnetic fields tend to stabilize the \ac{VSI} \citep{ng2013, ly2015, lp2018}, it was shown that the instability operates even in the presence of radiative diffusion \citep{sk2014} and non-ideal \ac{MHD} effects \citep{cb2020}. 

On the theoretical side, several studies have attempted to explore the linear-regime behavior of the VSI. \citet{ub1998} and \citet{u2003} employed a local (Boussinesq) approximation to study the VSI in a small section of an isothermal disk. They derived the local dispersion relation and provided the criteria along with the expressions for the maximum growth rate (see also \citealt{lp2018}). \citet{ng2013} extended the local analyses of \citet{ub1998} and \citet{u2003} to a semi-global analysis of an isothermal disk by allowing for the vertical variations of the background quantities. They found that the VSI consists of two distinct modes: (1) surface modes growing rapidly at small radial scales near the surfaces and (2) body modes growing relatively slowly at large radial scales in the main body of the disk.  \citet{bl2015} also analyzed the semi-global modes of the VSI in polytropic disks, finding that the surface modes grow fastest in the regions where the vertical shear is maximized. 

While these theoretical studies are useful for understanding the VSI operating in PPDs, most have been restricted to vertically isothermal disks. However, real \acp{PPD} are thermally stratified due to stellar irradiation \citep{cg1997, dc1998}, which preferentially heats the disk surfaces. Due to a large optical depth, the regions close to the midplane are protected against the stellar irradiation and thus remain cold. This naturally establishes a positive temperature gradient along the vertical direction. Since the vertical shear depends on the pressure gradient, the development of the \ac{VSI} and its nonlinear outcomes in thermally stratified disks  may be expected to differ from those in the isothermal counterparts. 
 
In this series of papers, we investigate the \ac{VSI} in thermally stratified disks using both linear stability analysis and hydrodynamic simulations. 
In this initial paper, we examine the impact of thermal stratification on the VSI through semi-global analyses. Our main objective here is to explore the effects of thermal disk stratification on the growth rates of surface and body modes. The findings of the present paper will be used to interpret the results of hydrodynamic simulations, which will be detailed in the subsequent paper.

This paper is organized as follows. In \autoref{sec:hydro}, we describe our models of thermally stratified disks and their properties. In \autoref{sec:local},  we briefly review the local theory of the \ac{VSI}.  In \autoref{sec:global}, we present a semi-global analysis of the VSI, assuming perturbations are local in the radial direction but global in the vertical direction, and compare the resulting eigenvalues and eigenfunctions for an isothermal disk with those for thermally stratified disks. Finally, we summarize and discuss our results in \autoref{sec:sum}.

\begin{figure*}
  \epsscale{1.0}
  \plotone{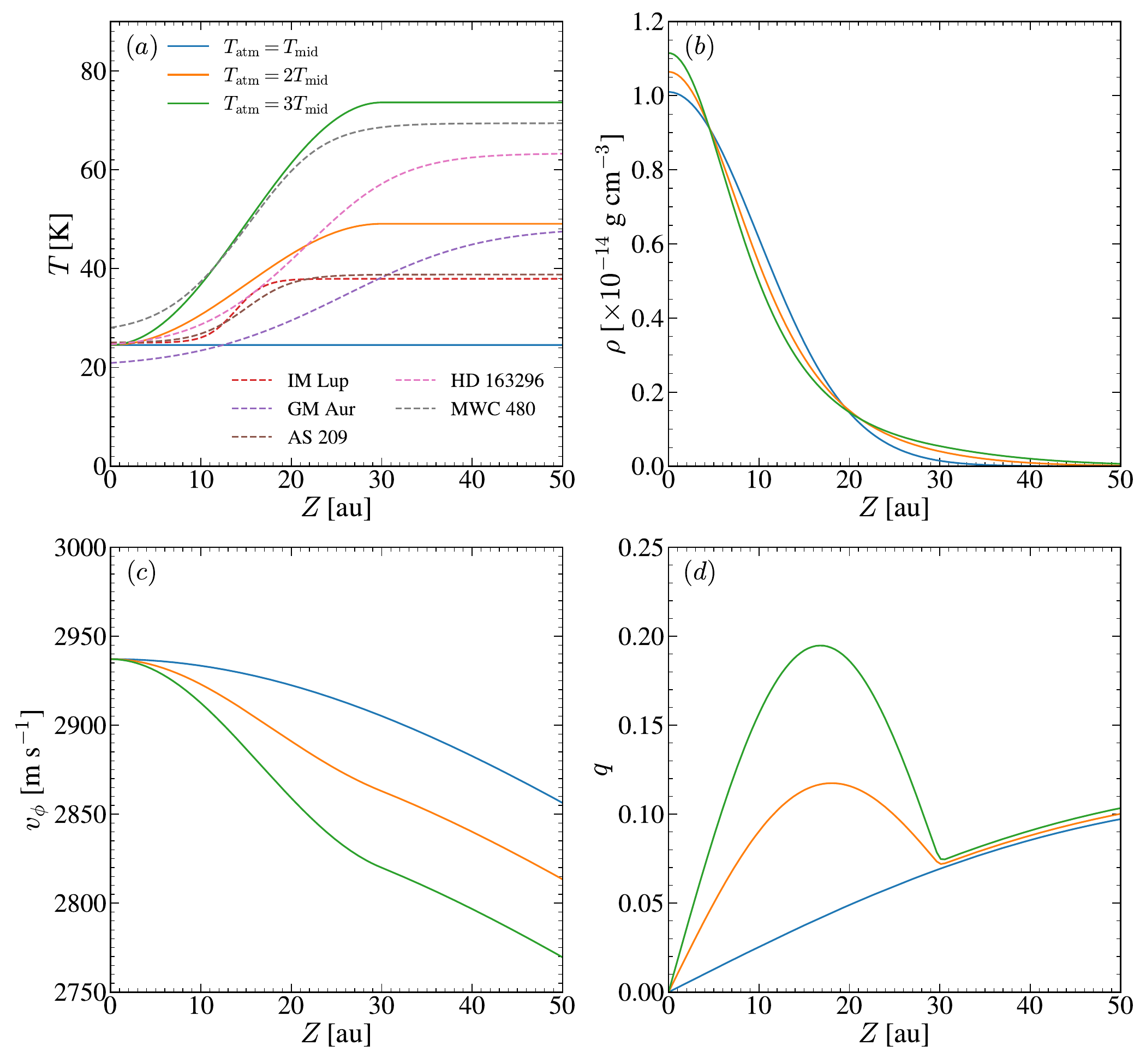}
  \caption{Vertical distributions of key quantities at the reference radius $R_0=100\unit{au}$ for disk models with differing thermal stratification ($n \equiv T_{\mathrm{atm}}/T_{\mathrm{mid}} = 1,2,3$):
  ($a$) the gas temperature $T$, ($b$) gas density $\rho$, $(c)$ rotational velocity $v_\phi=R\Omega$, and $(d)$ vertical shear $q = - R \partial \ln \Omega/\partial Z$.  In $(a)$, various dashed lines depict the temperature profiles for some observed \acp{PPD} from the MAPS survey \citep{lt2021}.
 }
  \label{fig:model}
\end{figure*}

\section{Disk Models} \label{sec:hydro}

In this study, we assume that disks are axisymmetric and utilize cylindrical coordinates $(R, Z)$. For the temperature distribution at the midplane, we adopt the model of \citet{bf2021},  $T_{\text{mid}}(R)=25.7(R/R_0)^{\alpha_T}\unit{K}$ at the midplane, where $R_0=100\unit{au}$ serves as the reference radius and $\alpha_T=-1/2$ represents the power-law slope of the temperature distribution. Following \citet{dd2003}, we allow the temperature to vary with $Z$ as
\begin{equation}
  T(R,Z)=
  \begin{cases}
  \displaystyle T_{\text{atm}}(R) + \left[T_{\text{mid}}(R)-T_{\text{atm}}(R)\right]\cos^2\left(\frac{\pi Z}{2Z_q}\right), \\
  \hfill \text{for  $Z < Z_q$}, \\
  T_{\text{atm}}(R), \hfill \text{for  $Z \geq Z_q$},
  \end{cases}
  \label{eq:temp}
\end{equation}
where $T_{\text{atm}} = n T_{\text{mid}}$ is the temperature in the disk ``atmosphere" located at $Z\geq Z_q$, with integer $n$ describing the temperature ratio. We consider a flared disk where the pressure scale height varies with $R$ as $H = 10(R/R_0)^{5/4}\unit{au}$. 
We in this work set $Z_q=3H$ and examine models with $n=1$, $2$, and $3$.

Our disks initially follow a power-law density distribution $\rho(R,0)=\rho_{0}(R/R_0)^{\alpha_{\rho}}$ in the midplane where $\alpha_{\rho}=-9/4$ being the slope of the density profile and $\rho_0=\rho(R_0,0)$: the corresponding surface density profile is $\Sigma = 2 H\rho(R,0)\propto R^{-1}$. The constant factor $\rho_{0}\approx 0.018M_* R_0^{-3}$, weakly depending on $n$, is chosen to ensure that the total mass of the disk equals 5\% of the mass of the central star $M_*=1\unit{M_\odot}$.\footnote{Since our models do not include self-gravity, the results of the present work and numerical simulations in the subsequent work are independent of the disk mass. However, the disk mass affects the optical depth of tracer elements, and consequently, synthetic ALMA images that we will produce in the next paper.}  The condition of hydrostatic equilibrium requires the density distribution along the vertical direction to obey
\begin{align}\label{eq:rho0}
  \rho(R,Z) =
  &\rho(R,0)\frac{c_s^2(R,0)}{c_s^2(R,Z)} \times \notag \\
  &\exp\left[-\int_0^{Z}\frac{1}{c_s^2(R,Z')}\frac{\partial\Phi}{\partial Z'}dZ'\right],
\end{align} 
where $c_s=(k_BT/\mu m_\text{H})^{1/2}$ is the isothermal speed of sound with the mean molecular weight $\mu=2.3$ and $\Phi\equiv -GM_*/(R^2+Z^2)^{1/2}$ is the gravitational potential due to the central star.

The angular velocity $\Omega$ of the disk rotation can be determined from the radial force balance
\begin{equation}\label{eq:eqOmg}
  \Omega^2(R,Z) = \Omega_K^2\sin\theta + \frac{1}{\rho r\sin^2\theta}\frac{\partial P}{\partial r},
\end{equation}
where $\Omega_K(R) = \sqrt{GM_*/R^3}$ is the Keplerian angular velocity, $P=c_s^2\rho$ is the gas pressure, $\theta=\cos^{-1}(Z/r)$ is the polar angle, and $r=R/\sin \theta$ is the spherical radius. We define the vertical shear parameter as 
\begin{equation}
q(R,Z) = - R\frac{\partial \ln \Omega}{\partial  Z}. 
\end{equation}
Note that $q(R,Z)=-q(R,-Z)$ for  a disk that is symmetric with respect to the midplane.

We explore three distinct disk models characterized by different values of $n=T_{\mathrm{atm}}/T_{\mathrm{mid}}$, namely, $n= 1$, 2, and 3. \cref{fig:model} illustrates the vertical profiles of various quantities at the reference radius $R=R_0$ for different $n$. In \cref{fig:model}($a$), the vertical temperature distributions of our models encompass a reasonable range of values typical of \acp{PPD} inferred from molecular line observations \citep{lt2021}.
As shown in \cref{fig:model}($c$), it is evident that the rotational velocity $v_\phi=R\Omega(R)$ declines more steeply with $Z$ for larger $n$. Consequently, the shear parameter $q$ increases with the thermal stratification in the regions with $Z<Z_q$.

Note that $q$ exhibits an approximately linear relationship with $Z$ in the isothermal disk, reaching $q = 0.10$ at $Z = 50\unit{au}$. However, in stratified disks, $q$ follows an approximately parabolic trend at $Z<30\unit{au}$ and linear behavior at higher $Z$. This results in peak values of $q=0.12$ and $0.19$ at $Z=18$ and $16\unit{au}$, for the $n=2$ and 3 disks, respectively. The height of the peak $q$ approximately corresponds to the region with the steepest change in disk temperature. We also note that $q$ is nearly independent of $n$ in the disk atmosphere with $Z\geq Z_q$ where $T$ remains constant. 
The density-weighted vertically averaged shear parameter is $\langle q\rangle=0.02$, $0.06$, and $0.10$ for disks with $n=1$, 2, and $3$, respectively. This is, of course, because a disk with a larger $n$ should rotate more slowly to compensate for the increased pressure gradient in \cref{eq:eqOmg}. 

\begin{figure}
    \epsscale{1.0}
    \plotone{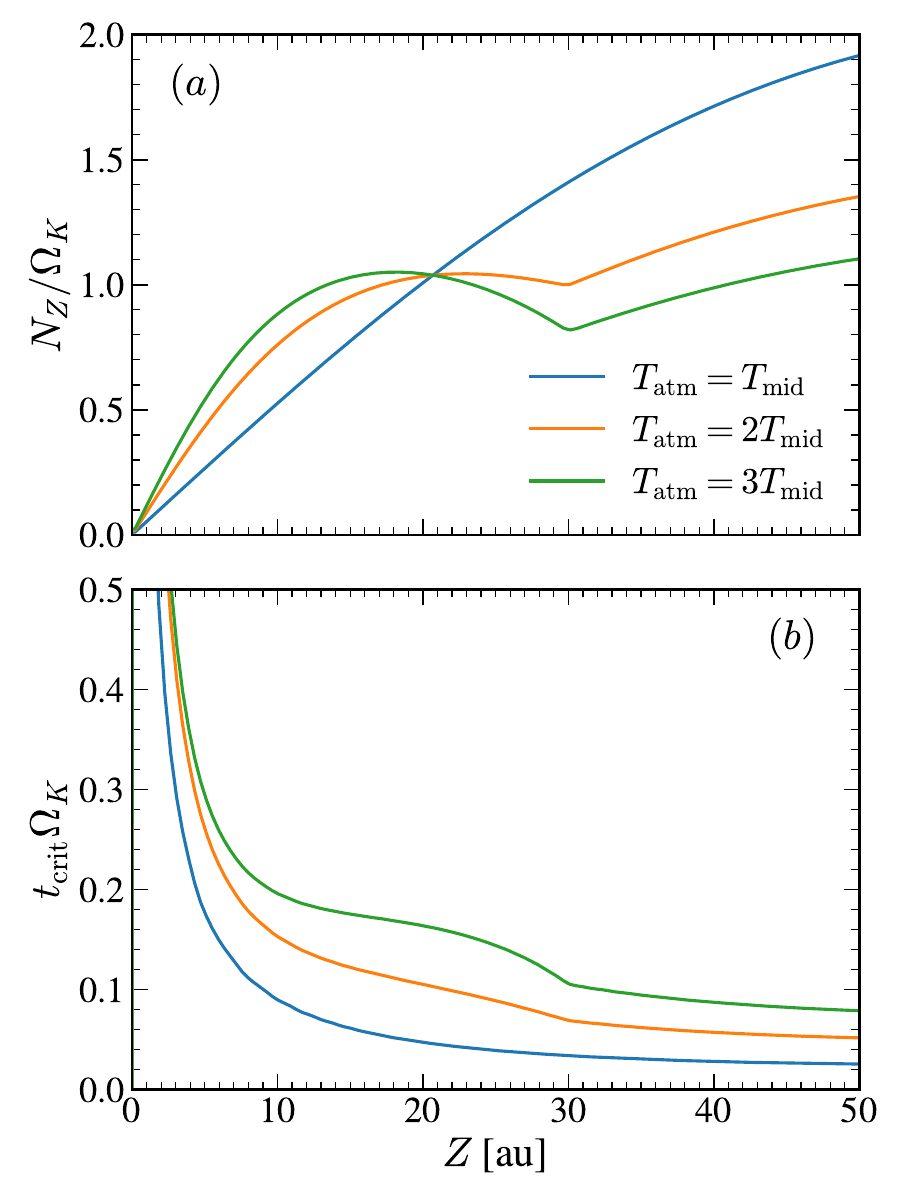}
    \caption{Vertical distributions of (a) the normalzied Brunt-V\"ais\"al\"a frequency $N_Z$  and (b) the critical cooling time $t_\text{crit}$ at the reference radius $R_0$ for the disks with $n=1$, $2$, and $3$.}
    \label{fig:critcool}
\end{figure}

In this paper, we adopt an isothermal equation of state by assuming very efficient cooling. This precludes the operation of buoyancy-induced vertical motions in our disks, which is known to significantly suppress the \ac{VSI} \citep{ng2013, ly2015}. When the cooling time $t_\text{cool}$ is finite, \citet{ly2015} showed that the criterion for the \ac{VSI} can be written as  
\begin{equation}
    t_\text{cool}  \lesssim t_\text{crit} \equiv \frac{|R\partial\Omega/\partial Z|}{N_Z^2},
\end{equation}
where 
\begin{equation}
N_Z^2 =  g\frac{\partial \ln(P^{1/\gamma}/\rho)}{\partial Z}  
\end{equation} 
is the Brunt-V\"ais\"al\"a frequency
with the adiabatic index $\gamma=1.4$ and the vertical gravity $g=\partial \Phi/\partial Z$ 
\citep[see also][]{u2003}. \Cref{fig:critcool} plots the vertical distributions of $N_Z/\Omega_K$ and $t_\text{crit}\Omega_K$ for disks with $n=1$, $2$, and $3$. It is apparent that thermal stratification enhances buoyancy at $Z\lesssim 20\unit{au}$ by creating a steeper density gradient (see \cref{fig:model}$b$), but reduces buoyancy at higher altitudes. However, the increased velocity shear extends the critical cooling time across all regions in thermally stratified disks. At low-latitude regions with $|Z|<20\unit{au}$ at $R=R_0$, an enhanced temperature stratification results in a larger vertical buoyancy as well as a larger vertical velocity shear. These two factors affect the development of the VSI in an opposing way: increased buoyancy stabilizes the \ac{VSI}, while greater velocity shear destabilizes it.  The combined effects of the two make a stratified disk a more favorable environment for the \ac{VSI} to develop.

\section{Local Theory} \label{sec:local}

 Here we provide a brief overview of the local theory of the \ac{VSI} to establish a foundation for the semi-global analysis presented in \autoref{sec:global}. As the \ac{VSI} excites inertial waves, which are essentially incompressible, the core dynamics of the \ac{VSI} can be captured by incompressible or anelastic models \citep{gs1967,ng2013, lp2018}. However, recent works have shown that gas compressibility \citep{ng2013, mp2015}, vertical buoyancy \citep{ly2015, lp2018}, magnetic fields \citep{lp2018, lk2022}, vertical dust stratification \citep{ly2017} and gas-dust drift \citep{ll2023} tend to stabilize the \ac{VSI}, although non-ideal MHD terms reduce the stabilizing effects of magnetic fields \citep{lk2022}.

For local, axisymmetric perturbations of the form $\propto\exp(st+ik_xx+ik_zz)$, where $s$ is the growth rate and $k_x$ and $k_z$ represent the radial and vertical wavenumber, respectively, in a rotating, shearing medium with uniform density, the dispersion relation of the \ac{VSI} is given by  
\begin{equation}\label{eq:idisp}
 s^2 =- \Omega^2 \frac{k_z^2}{k^2} \left(1+2q\frac{k_x}{k_z}\right),
\end{equation}
where $k=(k_x^2+k_z^2)^{1/2}$ \citep[e.g.,][]{lp2018}. The first term in the parentheses of \cref{eq:idisp} represents the frequency of inertial-gravity waves, indicating that the \ac{VSI} is inertial-gravity waves destabilized by vertical shear.
Note that the condition for the \ac{VSI} is
\begin{equation}\label{eq:cri}
 \frac{k_z}{k_x}<-2q.
\end{equation}
Since $q>0$ in the upper half of a PPD, the \ac{VSI}-prone modes should have $k_x/k_z<0$ there, while modes with $k_x/k_z>0$ can be unstable in the lower half of the disk, where $q<0$. From \Cref{eq:idisp}, it can be shown that the fastest-growing mode for fixed $k_z$ satisfies
\begin{equation}\label{eq:incomp}
    \left(\frac{k_x}{k_z}\right)^2 + \frac{1}{q}\left(\frac{k_x}{k_z}\right)-1 = 0.
\end{equation}
For $k_x/k_z\gg1$, \Cref{eq:incomp} yields an asymptotic solution 
\begin{equation}
\frac{k_x}{k_z}\approx -\frac{1}{q},
\end{equation} 
with the corresponding maximum growth rate of $s\approx \Omega|q|$ \citep{lp2018}.

\citet{lp2018} showed that the linear modes of the \ac{VSI} also manifest as non-linear solutions which are in turn subject to parasitic Kelvin–Helmholtz instability, driving turbulence with the maximum amplitude of $\delta v_{\mathrm{max}} \sim \Omega/k \sim c_s q$ (see also \citealt{cl2022}). 
Since the averaged shear parameter is  $\langle q\rangle =0.02$ and $0.10$ for $n=1$ and $3$, respectively, the local analysis predicts that the growth rate, radial wavelength, and turbulent amplitude of the \ac{VSI} in the thermally stratified disk with $n=3$ would be $\sim$5--15 times larger than those in the isothermal disk. Since the ratio of the velocity perturbations is
\begin{equation}
  \frac{v_x}{v_z} = -\frac{k_z}{k_x} \sim q,
  \label{eq:localratio}
\end{equation}
for the most unstable modes, the ratio of the radial to vertical kinetic energy in the $n=3$ disk would be higher than in the $n=1$ disk.

\section{Semi-global Analysis} \label{sec:global}

To properly account for the thermal stratification in the vertical direction, we adopt the quasi-global approach of \citet{ng2013} and \citet{bl2015} in which waves are treated as axisymmetric, local in the radial direction, and global in the vertical direction. In this section, we derive the perturbation equation for the thermally stratified disks, and solve it as a boundary-value problem to calculate eigenvalues and eigenfunctions.

\subsection{Perturbation Equation}\label{sec:ptb}

The basic equations of hydrodynamics governing isothermal axisymmetric flows in cylindrical coordinates $(R,\phi, Z)$ are as follows:
\begin{align}
   &\left(\frac{\partial}{\partial t} + {\bf V}\cdot \boldsymbol\nabla \right) \Pi  +  \frac{U}{R} +
  \frac{\partial  U}{\partial R} + \frac{\partial W}{\partial Z} = 0, \label{eq:ccon}\\
  &\left(\frac{\partial}{\partial t} + {\bf V}\cdot \boldsymbol\nabla \right)U - \frac{V^2}{R} = -
  c_s^2\frac{\partial\Pi}{\partial R}- \frac{\partial c_s^2}{\partial R} - \frac{\partial\Phi}{\partial R}, \label{eq:cmomR}\\
  &\left(\frac{\partial}{\partial t} + {\bf V}\cdot \boldsymbol\nabla \right)V + \frac{UV}{R} = 0, \label{eq:cmomphi}\\
  &\left(\frac{\partial}{\partial t} + {\bf V}\cdot \boldsymbol\nabla \right)W = -c_s^2\frac{\partial\Pi}{\partial Z} -\frac{\partial c_s^2}{\partial Z}-\frac{\partial\Phi}{\partial Z},\label{eq:cmomZ}
\end{align}
where $\Pi\equiv \ln \rho$, ${\bf V}=(U, V, W)$ is the velocity in the inertial frame,
${\bf V}\cdot \boldsymbol\nabla = U\partial/\partial R + W\partial/\partial Z$, and $\Phi=-GM_*/(R^2+Z^2)^{1/2}$ is the gravitational potential of the central star.

We use overbars to represent the initial background states.
\cref{eq:ccon,eq:cmomR,eq:cmomphi,eq:cmomZ} require that 
\begin{align}
  -\frac{\bar{V}^2}{R} &= -c_s^2\frac{\partial\bar{\Pi}}{\partial R} -\frac{\partial c_s^2}{\partial R} - \frac{\partial\Phi}{\partial R}, \label{eq:equil1} \\
  0 &= -c_s^2\frac{\partial\bar{\Pi}}{\partial Z}-\frac{\partial c_s^2}{\partial Z} - \frac{\partial \Phi}{\partial Z}, \label{eq:equil2}
\end{align}
and $\bar{U}=\bar{W}=0$ for an initial equilibrium. Note that \cref{eq:rho0,eq:eqOmg} are the formal solution of  \cref{eq:equil1,eq:equil2}.

We introduce dimensionless variables $\tilde{t} \equiv \Omega_0t$, $(r,z)\equiv  (R/R_0,Z/H_0)$, and $(u,v,w)\equiv (U,V,W)/c_0$, where $\Omega_0=\Omega(R_0,z=0)$, $c_0\equiv c_s(R_0,z=0)$, and $H_0\equiv c_0/\Omega_0$. 
Assuming that the perturbations have very small amplitudes, we linearize \cref{eq:ccon,eq:cmomR,eq:cmomphi,eq:cmomZ}  to obtain
\begin{align}
  &\frac{\partial\Pi'}{\partial\tilde{t}} = -h_0\left(\frac{u'}{r}+\frac{\partial u'}{\partial r}\right) - \frac{\partial w'}{\partial z} - w'\frac{\partial\bar{\Pi}}{\partial z}, \label{eq:conp} \\
  &\frac{\partial u'}{\partial\tilde{t}} = 2\frac{\bar{v}}{r}v' - h_0f(z)\frac{\partial\Pi'}{\partial r}, \label{eq:mup}\\
  &\frac{\partial v'}{\partial\tilde{t}} = -u'\left(\frac{\bar{v}}{r} + \frac{\partial\bar{v}}{\partial r}\right) - \frac{w'}{h_0}\frac{\partial\bar{v}}{\partial z}, \label{eq:mvp}\\
  &\frac{\partial w'}{\partial\tilde{t}} = -f(z)\frac{\partial\Pi'}{\partial z},   \label{eq:mwp}
\end{align}
where the primed quantities denote the perturbations, $h_0\equiv H_0/R_0 =0.1$ is the disk aspect ratio, $f(z)\equiv c_s^2(z)/c_0^2 =T(z)/T_\textrm{mid}$ represents the vertical dependence of the disk temperature, and $\bar{v}\equiv\bar{V}/R_0\Omega_0$.

\begin{figure*}
  \centering
  \plotone{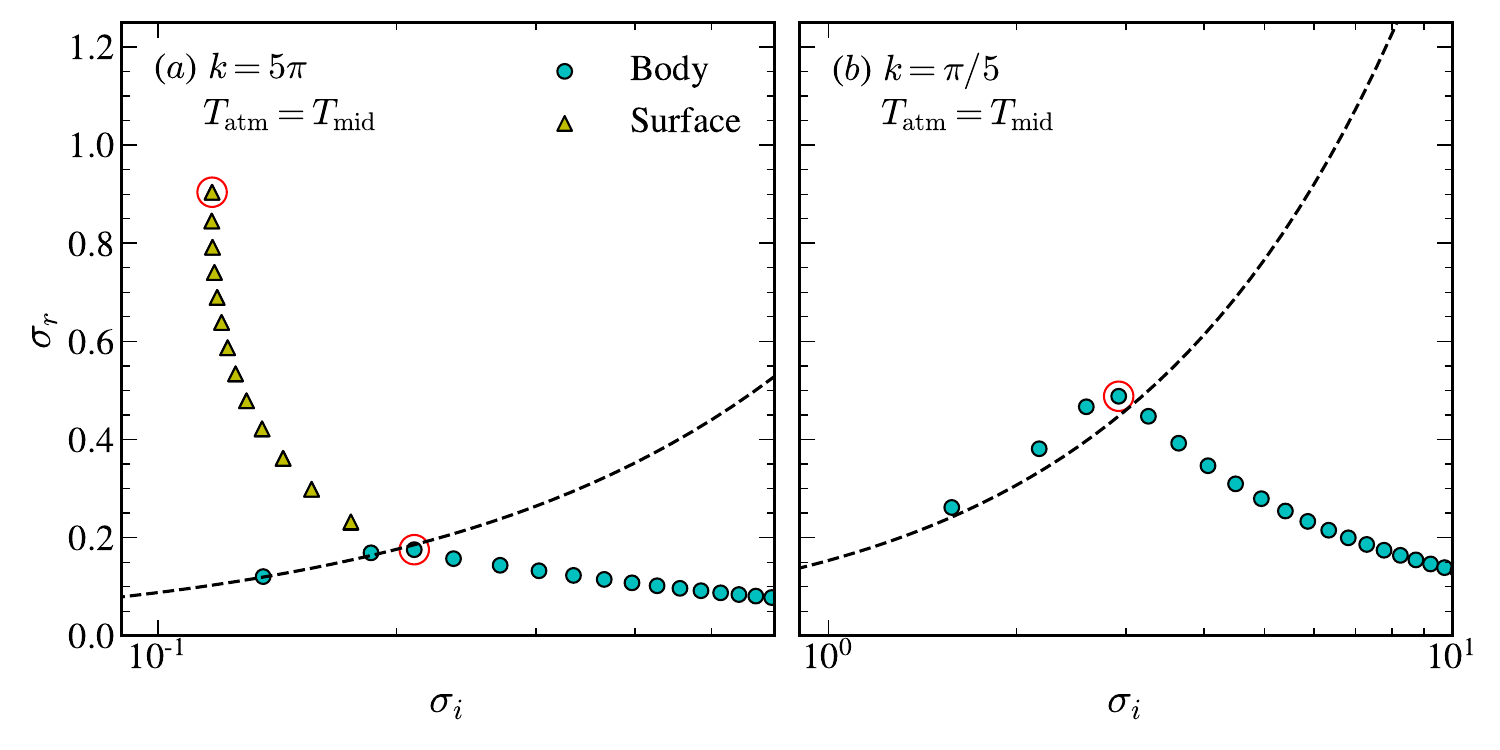}
  \caption{Growth rates ${\sigma}_r$ and oscillation frequencies ${\sigma}_i$ for the VSI modes for (${a}$) the radial wavenumber ${k} = 5\pi$ (or, $\lambda = 0.04H_0$) and ($b$) ${k}=\pi/5$ (or, $\lambda = H_0$) in the isothermal disk with $n=1$. Filled circles and triangles denote the body and surface modes, respectively.  Dashed lines represent \cref{eq:barker} for the location of the body modes in the limit of an infinitely extended disk. The red circles indicate the most unstable surface and body modes, whose eigenfunctions are plotted in \autoref{fig:Efun_iso}. 
   }\label{fig:Eval_iso}
\end{figure*}

Following \citet{ng2013}, we consider the regions around the reference radius $r=1$, and use the scaled radial and vertical distances as
\begin{equation}
\eta\equiv \frac{r-1}{h_0^2}\quad
\text{and}\quad
\zeta\equiv \frac{z}{h_0}.
\end{equation}
The different normalization is to ensure that the wavenumber in the $\eta$-direction is comparable to that in the $\zeta$-direction. In addition, we scale the perturbed density and velocities as 
\begin{equation}
\hat{\Pi}=\frac{\Pi'}{h_0},\quad 
\hat{u}=\frac{u'}{h_0}, \quad \hat{v}=v', \quad \hat{w}=w',
\end{equation}
since the VSI is incompressible and the perturbed radial velocity is much smaller than the other velocities. Finally, we rescale the time variable as 
${\tau}\equiv h_0\tilde{t}$ since the growth rates of the VSI are roughly proportional to $h_0$ \citep{u2003,au2004}. Applying these scaled variables to \cref{eq:conp,eq:mup,eq:mvp,eq:mup,eq:mwp} and dropping the terms proportional to $h_0^2$, we obtain
\begin{align}
  & 0 = \frac{\partial\hat{u}}{\partial {\eta}} + \frac{\partial \hat{w}}{\partial {\zeta}} + \hat{w}\frac{\partial\bar{\Pi}}{\partial {\zeta}},\label{eq:pert_pi} \\
 & 0 = 2\hat{v}-f({\zeta})\frac{\partial\hat{\Pi}}{\partial {\eta}},\label{eq:pert_u} \\
  & \frac{\partial \hat{v}}{\partial{\tau}} = -\frac{\hat{u}}{2}+\hat{w}\bar{q}({\zeta}), \label{eq:pert_v} \\
 & \frac{\partial \hat{w}}{\partial{\tau}} = -f({\zeta})\frac{\partial\hat{\Pi}}{\partial {\zeta}}, \label{eq:pert_w}
\end{align}
where $\bar{q}\equiv-h_0^{-2}{\partial\bar{v}}/{\partial {\zeta}}$. Note that \cref{eq:pert_pi,eq:pert_u,eq:pert_v,eq:pert_w} reduce to Equations (A15) to (A18) of \citet{ng2013} for a vertically isothermal disk under the condition that $f({\zeta})=1$, $\partial\bar{\Pi}/\partial {\zeta}=-{\zeta}$, and $\bar{q} = -(\alpha_T/2){\zeta}$.

We consider the perturbations of the form
\begin{equation}
\begin{pmatrix}
\hat{\Pi} \\ \hat{u} \\ \hat{v} \\ \hat{w}
\end{pmatrix}
({\eta},{\zeta},{\tau})
=
\begin{pmatrix}
\hat{\Pi}({\zeta}) \\ \hat{u}({\zeta}) \\ \hat{v}({\zeta}) \\ \hat{w}({\zeta})
\end{pmatrix}
\exp(i{k}{\eta}+{\sigma}{\tau}),
\end{equation}
where ${\sigma} \,{ (=s/[h_0\Omega_0])}$  and $k\, {(= h_0H_0k_R)}$ denote the dimensionless frequency and radial wavenumber, respectively, for the dimensional frequency $s$ and radial wavenumber $k_R$. Then, \cref{eq:pert_pi,eq:pert_u,eq:pert_v,eq:pert_w} can then be combined to yield 
the perturbation equation
\begin{equation}
  \frac{\partial^2\hat{\Pi}}{\partial {\zeta}^2} + \left[\frac{\partial\bar{\Pi}}{\partial {\zeta}}+\frac{\partial\ln f}{\partial {\zeta}} + 2i{k}\bar{q}\right]\frac{\partial\hat{\Pi}}{\partial {\zeta}} - {\sigma}^2{k}^2\hat{\Pi} = 0. \label{eq:radial1}
\end{equation}
For a vertically isothermal disk with $f({\zeta})=1$, $\partial\bar{\Pi}/\partial {\zeta}=-{\zeta}$, and $\bar{q} = -(\alpha_T/2){\zeta}$, \cref{eq:radial1} again agrees with Equation (39) of \citet{ng2013} or Equation (28) of \citet{bl2015}. A different degree of thermal  stratification alters $f({\zeta})$, $\bar \Pi (\zeta)$ and $\bar{q}$. 

\Cref{eq:radial1} can be solved for eigenvalues ${\sigma}$ for a given wavenumber ${k}$ subject to boundary conditions. We impose the no-flow boundary conditions, $\hat{w}=0$, which is equivalent to $\partial \hat{\Pi}/\partial {\zeta}=0$ at the vertical boundaries, as derived from \cref{eq:pert_w}, located at $\zeta=\pm5$. The resulting eigenvalues are generally complex numbers, indicating that the semi-global VSI is in fact overstability. The imaginary part, ${\sigma}_i$, of the eigenvalues represents the oscillation frequency, while the real part, ${\sigma}_r$, gives the growth or damping rate depending on its sign.

\begin{figure*}
  \centering
  \plotone{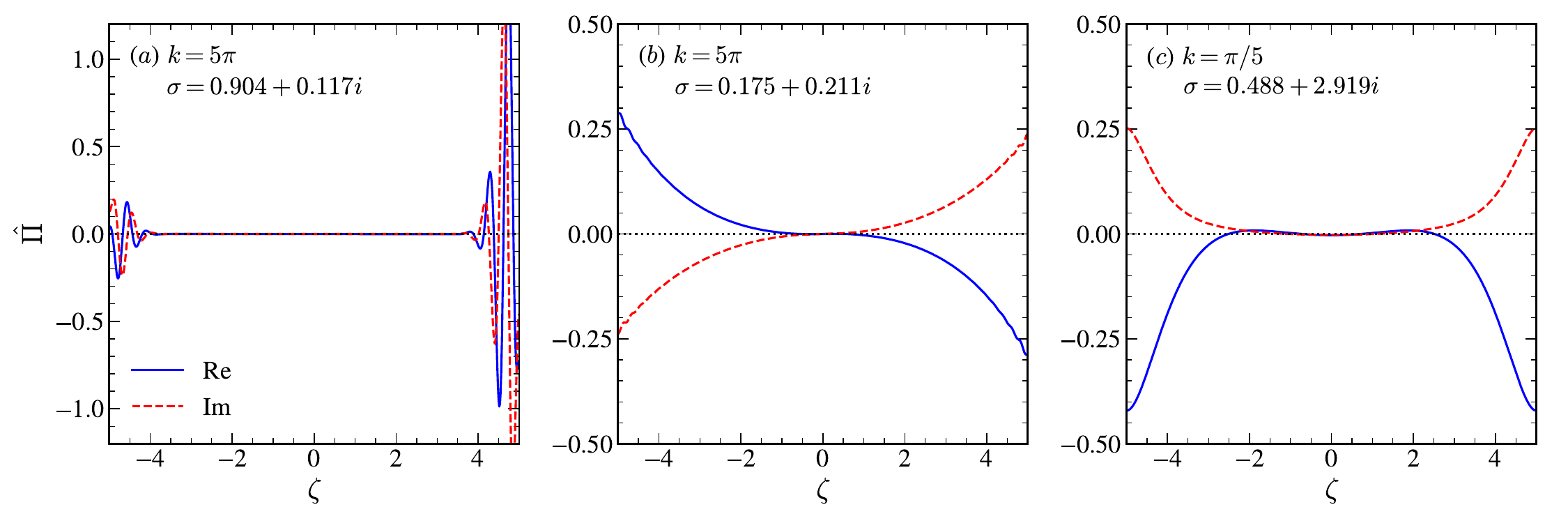}
  \caption{Eigenfunctions of the most unstable ($a$) surface mode with $k=5\pi$, ($b$) body mode with $k=5\pi$, and ($c$) body mode with $k=\pi/5$ in the $n=1$ disk. The blue and red lines correspond to the real and imaginary parts of the eigenfunctions, respectively. Each eigenfunction is normalized such that $\int|\hat{\Pi}|d\zeta = 1$. 
   }\label{fig:Efun_iso}
\end{figure*}

\subsection{Eigenvalues and Eigenfunctions}\label{sec:eig}

We numerically solve \cref{eq:radial1} using $\mathtt{DEDALUS}$, a pseudo-spectral code developed by \citet{bv2020}, specifically designed for solving partial differential equations including linear eigenvalue problems. $\mathtt{DEDALUS}$ employs a Chebyshev collocation method on $N$ Gauss-Lobatto grid points, facilitating the transformation of \cref{eq:radial1} into a matrix eigenvalue problem solvable by the QZ algorithm \citep{gl1996, b2001}. We ensure the convergence of the eigenvalues by comparing the results with $N=400$ and $600$ grid points.

\subsubsection{Vertically Isothermal Disk}

\Cref{fig:Eval_iso} plots the resulting discrete eigenvalues for the \ac{VSI} with $k=5\pi$ and $k=\pi/5$, corresponding to the dimensional radial wavelength $\lambda_R=2\pi/k_R=0.04H_0$ and $1H_0$, respectively, in the $n=1$ disk. The ordinate and abscissa give the real part ${\sigma}_r$ and imaginary part ${\sigma}_i$ of the eigenvalues, respectively. \Cref{fig:Efun_iso} plots the eigenfunctions for $\hat{\Pi}$ of the modes marked by the red circles in \Cref{fig:Eval_iso}. 

We classify the unstable modes into surface and body modes based on the vertical dependence of $\hat{\Pi}$. For the surface modes, the eigenfunctions exhibit significant variation near the vertical boundaries and diminish near the midplane.  These modes arise in the regions where the vertical shear is strongest, and are thus analogous to the localized non-oscillatory \ac{VSI} disturbances described in \autoref{sec:local}, modified by the imposed boundary conditions.
In contrast, the eigenfunctions of the body modes maintain a finite amplitude throughout the main body of the disk. These modes are semi-global inertial-gravity waves destabilized by the vertical shear \citep{ng2013,bl2015}.

When $k=5\pi$, \Cref{fig:Eval_iso} shows that the isothermal disk possesses 13 surface modes and a number of body modes. These surface modes are further categorized into symmetric and antisymmetric modes with respect to the midplane, with equal eigenvalues.
In the isothermal disk, the growth rate of the surface modes increases as the oscillation frequency $\sigma_i$ decreases. The eigenfunction of the most unstable surface mode depicted in \Cref{fig:Efun_iso}($a$) corresponds to an antisymmetric mode. The surface modes tend to oscillate more slowly near the surface as $\sigma_i$ increases.

\begin{figure*}
  \centering
  \plotone{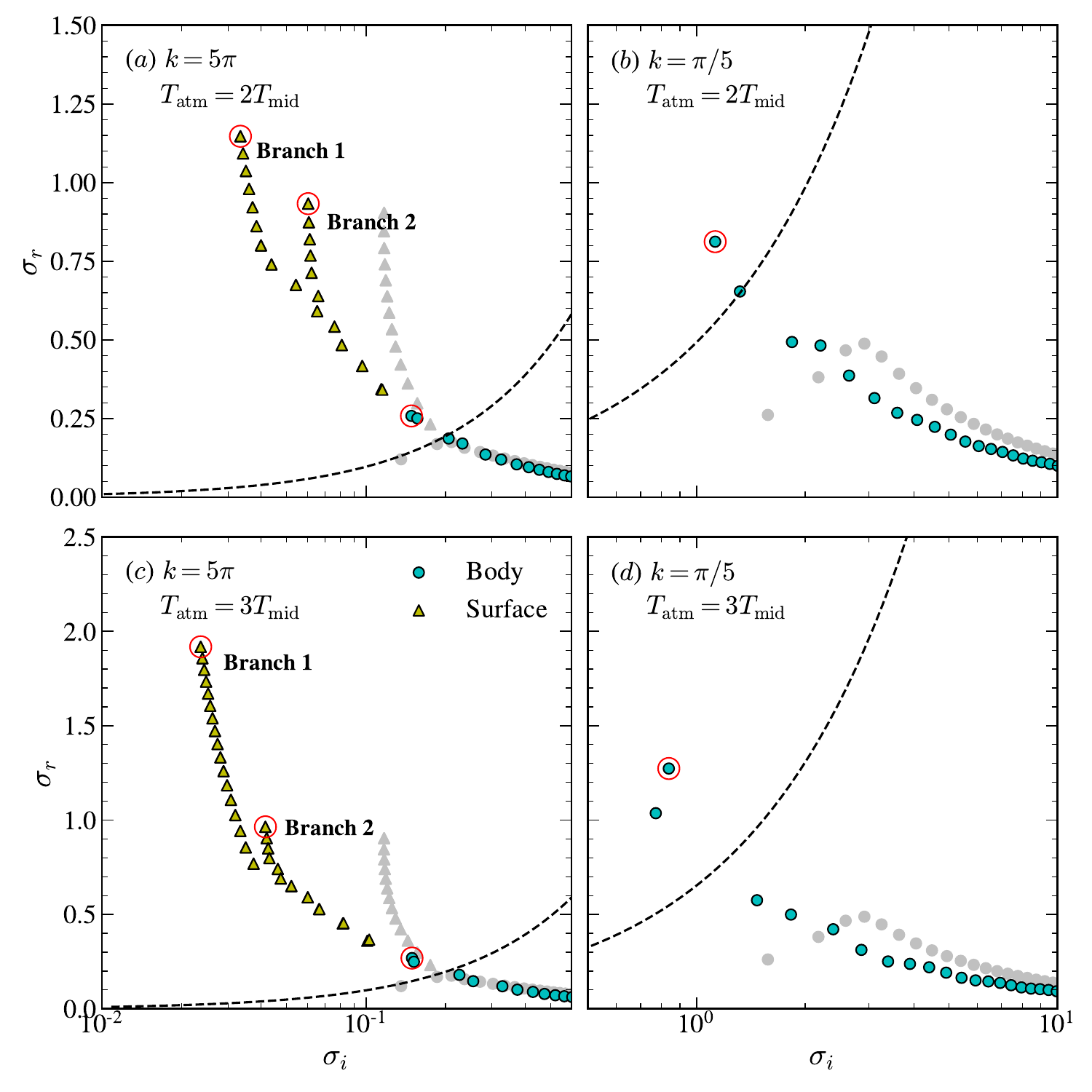}
  \caption{Growth rates of the unstable modes \emph{vs.} their oscillation frequencies for the thermally stratified disks with (upper) $n=2$ and (lower) $n=3$. The left and right panels correspond to the cases with $k=5\pi$ and $k=\pi/5$, respectively.  Dashed lines represent \cref{eq:barker} for the location of the body modes in the case of an infinitely extended isothermal disk.
  Filled circles and triangles denote the body and surface modes, respectively. The red circles mark the most unstable surface and body modes, whose eigenfunctions are illustrated in \autoref{fig:Efun_str}. 
  For comparison, the results of the isothermal disk with $n=1$ are plotted as grey markers. 
  \label{fig:Eval_str}}
\end{figure*}

\begin{figure*}
  \centering
  \epsscale{1.15}
  \plotone{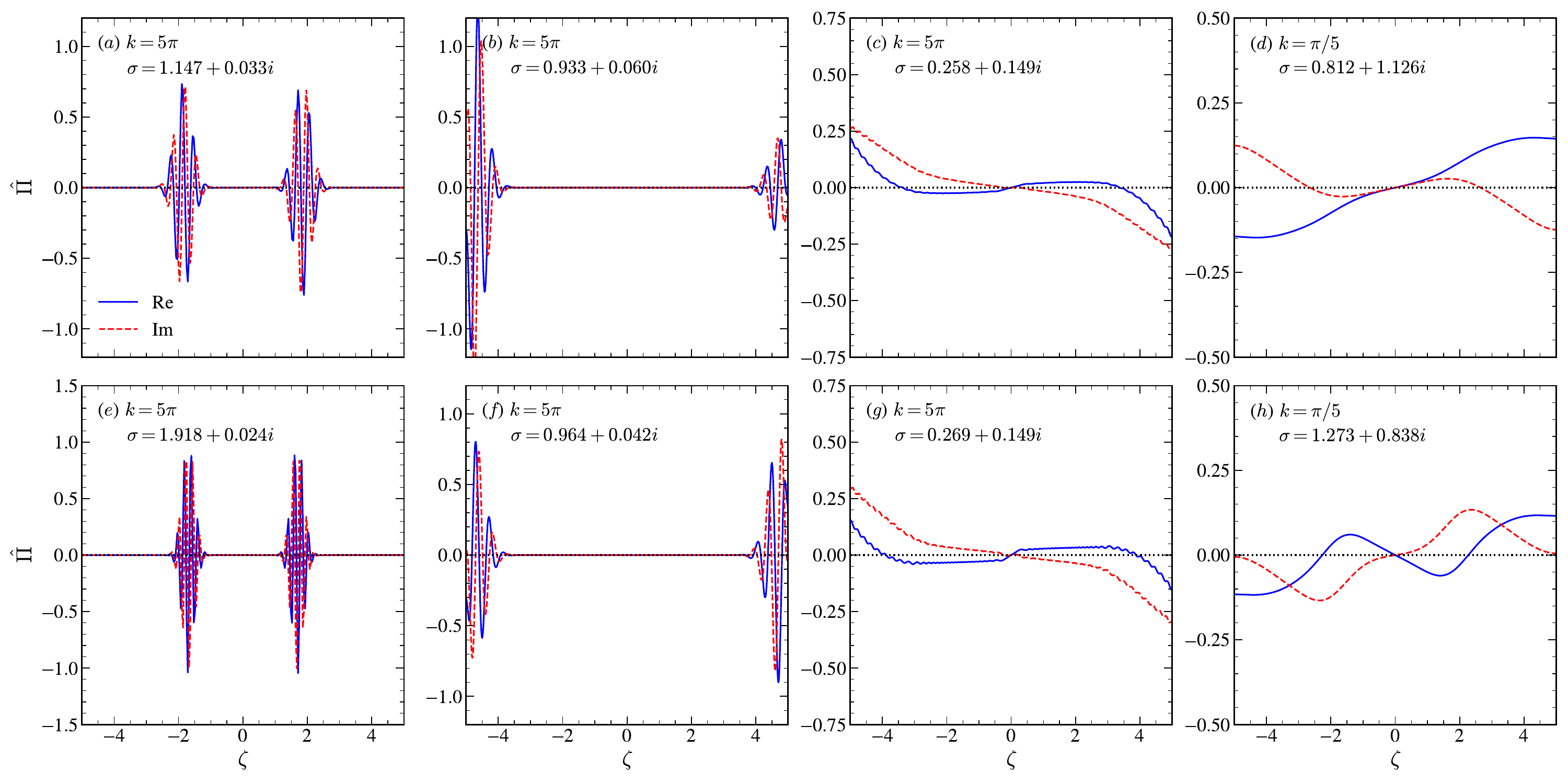}
  \caption{Eigenfunctions of the most unstable (\textit{a,e}) first-branch surface mode with $k=5\pi$, (\textit{b,f}) second-branch surface mode with $k=5\pi$,
  (\textit{c,g}) body mode with $k=5\pi$, and (\textit{d,h}) body mode
 with $k=\pi/5$. The upper and lower panels are for the thermally stratified disks with $n = 2$ and 3, respectively. The blue and red lines correspond to the real and imaginary parts of the eigenfunctions, respectively.  Each eigenfunction is normalized such that $\int|\hat{\Pi}|d\zeta = 1$. 
 }
  \label{fig:Efun_str}
\end{figure*}

Body modes in general exhibit a lower growth rate compared to surface modes.  \citet{bl2015} demonstrated that
for a vertically infinite, isothermal disk with a Gaussian density distribution (i.e., $\rho\propto e^{-\zeta^2/2}$), the body modes have analytic eigenfrequencies
\begin{equation}\label{eq:barker}
  \sigma = \pm\frac{\sqrt{m}}{ik}\left(1 - 2ik \left.\frac{d\bar{q}}{d\zeta}\right|_{\zeta=0}\right)^{1/2},
\end{equation}
where $m$ is an integer and the term $d\bar{q}/d\zeta|_{\zeta=0}$ arises from the radial temperature gradient of the disk. The dashed lines in \Cref{fig:Eval_iso} represent \cref{eq:barker} as a continuum. For $k=5\pi$, the three body modes with the smallest $\sigma_i$ values match well with the prediction of \cref{eq:barker}. Body modes can also be classified as symmetric and antisymmetric; however, unlike surface modes, they have different eigenfrequencies.\footnote{The symmetric and antisymmetric  body modes are referred to as breathing and corrugation modes, respectively, in \citet{ng2013}.} The two body modes with the smallest $\sigma_i$ correspond to the fundamental antisymmetric and symmetric modes \citep{ng2013}. The third body mode, which is most unstable for $k=5\pi$, represents the first overtone antisymmetric mode, with its eigenfunction shown in \Cref{fig:Efun_iso}($b$). The other body modes, which have a smaller growth rate with increasing $\sigma_r$, deviate from the behavior predicted by \cref{eq:barker} due to the truncated vertical boundaries in our model disks.

We find that the surface modes disappear when the radial wavenumber $k$ surpasses the threshold $k_\mathrm{crit} \sim \pi/2$, as illustrated in \cref{fig:Eval_iso} for $k=\pi/5$.
This can be understood qualitatively through \cref{eq:cri}. Local unstable modes typically exhibit the vertical wavelength 
\begin{equation}
\frac{\lambda_z}{H_0} = \frac{\pi h_0}{k q} \sim 2.5,
\end{equation}
which is too large to be confined near the surfaces: modes with sufficiently large $\lambda_z$ transform into body modes that extend globally in the $z$-direction \citep[see also][]{bl2015}. This global effect also makes the eigenvalues at low frequencies deviate slightly from the prediction of \cref{eq:barker}. The most unstable body mode in this case is the first overtone symmetric modes, plotted in \Cref{fig:Efun_iso}($c$). The body modes tend to oscillate more rapidly as $\sigma_i$ increases.

\subsubsection{Thermally Stratified Disks}

\Cref{fig:Eval_str} plots the discrete eigenfrequencies of the \ac{VSI} with $k=5\pi$ and $\pi/5$ in the thermally stratified disks with (upper) $n=2$ and (lower) $n=3$.  
\Cref{fig:Efun_str} plots the eigenfunctions of the most unstable surface and body modes marked by the red circles in \Cref{fig:Eval_str}.

In comparison with the isothermal disk, thermal stratification induces several changes. Firstly, the surface modes bifurcate into two branches due to the non-monotonic distribution of the shear rate, as shown in \autoref{fig:model}($d$). The first branch, situated on the smaller $\sigma_i$ side, is associated with the maximum shear at $Z=16\unit{au}$, resulting from the temperature gradient. 
In contrast, the second branch, positioned on the relatively higher $\sigma_i$ side, is related to large shear at the disk surfaces, akin to the $n=1$ disk. The eigenfunctions of the most unstable surface modes, shown in \cref{fig:Efun_str}, clearly reveal that the first and second branches of the surface modes involve perturbations localized at $|\zeta|\sim 1.6$ and 5, respectively. 

Secondly, the surface modes in the thermally stratified disks exhibit larger growth rates, by a factor of $\sim1.4$ and 2.1 for the $n=2$ and 3 disks, respectively, when $k=5\pi$, compared to the isothermal disk. This increase is attributed to the enhanced vertical shear in the stratified disks. However, the growth rate of the most unstable surface mode in the second branch shows only a slight increase, by a factor of $\sim1.03$ and 1.07 compared to the isothermal disk. This is expected since the shear at the disk boundary is almost independent of $n$. Thirdly, \cref{eq:barker} does not accurately predict the location of the body modes in the stratified disks primarily due to significant variations of $\bar{q}$ with respect to $\zeta$. When $k=5\pi$, there are two body modes with $\sigma_r$ exceeding the prediction of \cref{eq:barker}. When $k=\pi/5$, in contrast, the disks exclusively host body modes whose growth rates increase with the degree of thermal stratification owing to enhanced shear. The critical radial wavenumber is $k_\textrm{crit}\sim\pi$ and $2\pi/3$ for the $n=2$ and 3 disks, respectively. In both the $k=5\pi$ and $\pi/5$ cases, the most unstable body mode manifests as an antisymmetric mode. 

\subsubsection{Energy Ratio}

\begin{figure*}
  \centering
  \epsscale{1.15}
  \plotone{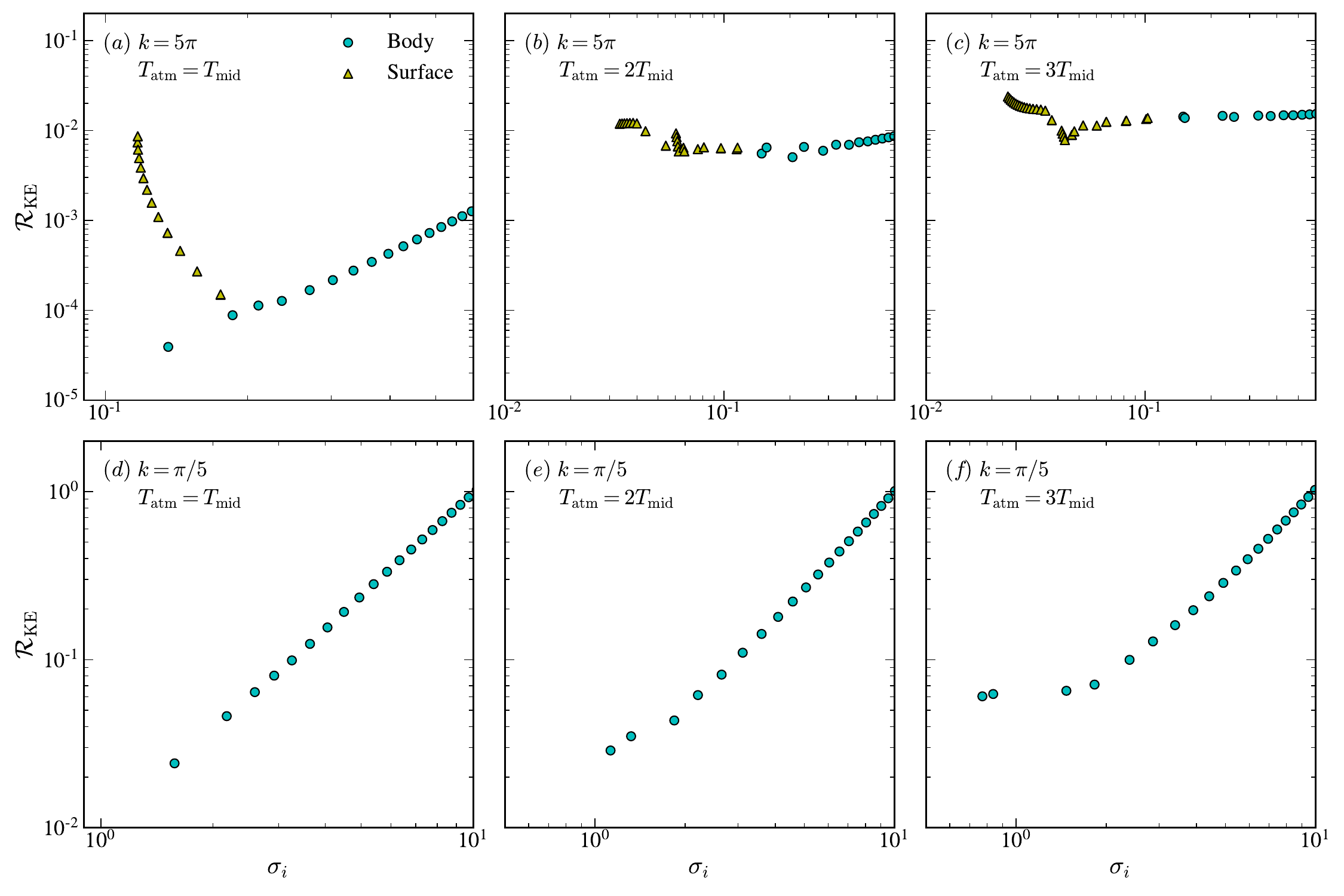}
  \caption{
  The ratio $\mathcal{R}_\mathrm{KE}$ of the kinetic energy densities in the radial and vertical directions for the modes shown in \Cref{fig:Eval_iso,fig:Eval_str}. The top and bottom panels correspond to the modes with radial wavenumbers $k=5\pi$ and $\pi/5$, respectively. Each column represents the $n=1$, 2, and 3 disk, from left to right. The energy ratio tends to be larger for smaller $k$ and larger $n$.
}
  \label{fig:globalratio}
\end{figure*}

\citet{ng2013} showed that in an isothermal disk, velocity perturbations induced by the \ac{VSI} are dominated primarily by the vertical component. But, how does the ratio of the radial to vertical kinetic energy vary with the degree of thermal stratification. This ratio affects the observability of \ac{VSI}-induced perturbations in stratified disks with modest inclinations \citep{bf2021}, which we will explore by using numerical simulations in the subsequent paper. Here, we present the energy ratio prediction from our semi-global analysis.

Once we obtain the eigenvalues $\sigma$ and the related eigenfunctions $\hat{\Pi}$, we use \cref{eq:pert_pi,eq:pert_w,eq:radial1} to calculate the perturbed radial and vertical velocities as 
\begin{align}
  \hat{u} &= 
  -f(\zeta)\left(i\sigma k\hat{\Pi} + \frac{2\bar{q}}{\sigma}\frac{\partial\hat{\Pi}}{\partial\zeta}\right)
  \label{eq:global_u} \\
  \hat{w} &= -\frac{f(\zeta)}{{\sigma}}\frac{\partial\hat{\Pi}}{\partial \zeta}, \label{eq:global_w}
\end{align}
respectively. 
Then, the ratio of the vertically-averaged kinetic energies in the radial and vertical directions is given by 
\begin{align}
    \mathcal{R}_\mathrm{KE}\equiv   \frac{\displaystyle\int  \frac{1}{2} \rho|h_0\hat{u}|^2d\zeta}
    { \displaystyle\int \frac{1}{2} \rho|\hat{w}|^2d\zeta}.
\end{align}

\autoref{fig:globalratio} plots the ratio $\mathcal{R}_\mathrm{KE}$ for the modes shown in \Cref{fig:Eval_iso,fig:Eval_str}. The upper and lower panels correspond to radial wavelengths of $k=5\pi$ and $k=\pi/5$, respectively. Each column represents the disks with $n=1$, 2, and 3 from left to right. Overall, $\mathcal{R}_\mathrm{KE}$ is larger in more thermally stratified disks and for modes with smaller $k$. It tends to decrease with increasing $\sigma_i$ for the surface modes and increases for the body modes. This behavior can be qualitatively understood from \cref{eq:localratio}, which suggests that stronger shear induces more kinetic energy in the radial direction. Surface (body) modes with higher $\sigma_i$ exhibit less (more) rapid vertical variations of the eigenfunction, corresponding to lower (higher) $k_z$. For fixed $k_z$, modes with smaller $|k_x|$ have larger $|v_x'/v_z'|$, and hence, larger $\mathcal{R}_\mathrm{KE}$. When $k=\pi/5$, the fundamental mode with smallest $\sigma_i$ in the $n=3$ disk has $\mathcal{R}_\mathrm{KE}$ that is larger by a factor of $\sim3$ compared to that in the $n=1$ disk.

 \section{Summary and Dicussion} \label{sec:sum}

In this paper, we analyze the linear stability of the \ac{VSI} in thermally stratified disks using semi-global approaches. To model realistic disks, where the atmospheric temperature $T_\textrm{atm}$ is higher than the midplane temperature $T_\textrm{mid}$ due to stellar irradiation, we consider three types of disks characterized by the ratios $n=T_\textrm{atm}/T_\textrm{mid}=1, 2$, and 3
(see \autoref{eq:temp}). Hydrostatic equilibrium dictates that a disk with higher $n$ exhibits a more rapid decrease in density and rotational angular velocity $\Omega$ with increasing $Z$, leading to a higher vertical shear rate, defined as $q=-R\partial \ln \Omega/\partial Z$.

In the isothermal disk with $n=1$, $q$ increases almost linearly with $Z$, reaching $q=0.10$ at $Z/H=5$. In contrast, in the thermally stratified disks, $q$ is approximately represented by a parabolic function at $Z/H<3$ and a linear function at higher $Z$, resulting in maximum values $q=0.12$ and $0.19$ at $Z/H=1.8$ and $1.6$ for the $n=2$ and 3 disks, respectively (see \cref{fig:model}$d$). The local theory predicts that disks with greater thermal stratification is more prone to the \ac{VSI} and have a larger radial wavelength and a larger ratio of the radial to vertical turbulence kinetic energy compared to the isothermal counterpart.

We impose small-amplitude axisymmetric perturbations to the background equilibrium state. We assume the perturbations to be semi-global, that is, the typical legnth scale of perturbations is much shorter than (comparable to) the scale over which disk properties change in the radial (vertical) direction. Using the reduced model of \citet{ng2013}, we derive the perturbation equation relevant to the thermally stratified disk (see \autoref{eq:radial1}). By solving the perturbation equation subject to no-flow boundary conditions, we calculate the discrete eigenvalues and associated eigenfunctions. 

For the isothermal disk, our semi-global analysis recovers the results of \citet{ng2013} and \citet{bl2015}.
For example, the \ac{VSI} is categorized into surface and body modes: surface modes are analogous to the local \ac{VSI} occurring predominantly near the disk surfaces, while body modes are semi-global inertial-gravity waves with perturbations distributed over the main body of the disk, encompassing the midplane. The surface modes generally have larger growth rates than the body modes. The surface modes are discernible only when the radial wavenumber $k_R$ exceeds the threshold value $k_\textrm{crit} \sim \pi/2$, and disappear for small $k_R$, where perturbations fail to localize effectively near the disk surfaces.

Thermal stratification not only amplifies the growth rates of both surface and body modes, but also increases the \ac{VSI}-driven kinetic energy in the radial direction relative to that in the vertical direction. Moreover, the non-monotonic distribution of the shear rate leads to the bifurcation of surface modes into two branches: the first branch, associated with the strongest shear around $Z\sim16\unit{au}$, at the reference radius $R_0=100\unit{au}$ in our models, exhibits a higher growth rate compared to the second branch corresponding to the strong shear at the disk surfaces ($Z=50\unit{au}$). The surface modes exist only when the wavenumber is larger than $k_\text{crit} \sim \pi$ and $2\pi/3$ in the $n = 2$ and 3 disks, respectively.

Regardless of thermal stratification, surface modes consistently exhibit larger growth rates than body modes, existing only when the radial wavenumber $k_R$ exceeds a certain threshold. Conversely, body modes are present across all $k_R$, with larger growth rates for smaller $k_R$. This suggests that numerical simulations of the \ac{VSI} will first capture the most unstable surface modes with sufficiently large $k_R$ permitted by resolution \citep{ng2013, sk2014}. As the perturbations, initially concentrated in regions with the strongest shear, propagate toward the midplane, the body modes with smaller $k_R$ will be excited and subsequently amplify, inducing turbulence in the disk. This transition occurs earlier in disks with higher $n$, as will be demonstrated in the forthcoming paper.

 In this paper we neglect the stabilizing effect of vertical buoyancy by taking an isothermal equation of state, which can be justified only when the cooling time is sufficiently short \citep[see e.g.,][]{ng2013,ly2015, lp2018}. While thermal stratification increases the critical cooling timescale needed for the onset of the \ac{VSI} (see \cref{fig:critcool}), the isothermal assumption may not be valid at large radii in \acp{PPD}, where the gas cooling can be inefficient. For example, \citet{pb2023} found that dust coagulation, which occurs when the fragmentation velocity is sufficiently high, can shrink the size of \ac{VSI}-active regions by making the disk cooling inefficient at large radii \citep[see also][]{fo2021}.

\citet{bf2021} demonstrated that the turbulence induced by the \ac{VSI} in a vertically isothermal disk is observable through ALMA molecular line observations. Given the significant role turbulence plays in affecting planet formation by regulating dust distribution and particle growth \citep{lu2019}, it becomes imperative to investigate the characteristics of \ac{VSI}-driven turbulence in more realistic disk models. In the subsequent paper, we will present results from direct hydrodynamic simulations aimed at elucidating the enhanced \ac{VSI} in thermally stratified disks. 
 
\section*{Acknowledgments}

We are grateful to the referee for detailed and 
constructive comments. 
The work of H.-G.Y.\ was supported by the grants of Basic Science Research Program through the National Research Foundation of Korea (NRF) funded by the Ministry of Education (2022R1A6A3A13072598). The work of W.-T.K.\ was supported by grants of the National Research Foundation of Korea (2020R1A4A2002885 and 2022R1A2C1004810). 
JB acknowledges support from NASA XRP grant No. 80NSSC23K1312.

\bibliographystyle{aasjournal}
\bibliography{theory}

\end{document}